\documentstyle[12pt]{article}
\title{\bf Poisson-Lie T-Duality and Bianchi Type Algebras}
\vspace{20mm}
\author{M. A. Jafarizadeh$^{a,b}$ \thanks{E-mail: jafarzadeh@ark.tabrizu.ac.ir} , 
A. Rezaei-Aghdam$^{a}$ \thanks{E-mail: a-rezaei-a@ark.tabrizu.ac.ir} \\ 
\\
\\
$^a${\small Department of Theoretical Physics, Tabriz University, Tabriz  51664, Iran.} \\   
$^b${\small Institute for Studies in Theoretical Physics and Mathematics, Tehran, 19395-1795, Iran.}} 
\begin{document}
\maketitle
\vspace{10mm}
\begin{abstract}
All Bianchi bialgebras have been obtained. By introducing a non-degenerate
adjoint invariant inner product over these bialgebras the associated Drinfeld
doubles have been constructed, then by calculating the coupling matrices for
these bialgebras several $\sigma$-models with Poisson-Lie symmetry have been obtained.
Two simple examples as prototypes of Poisson-Lie dual models have been given.

\end{abstract}
\newpage
\section{Introduction}

$\; \; \; \;$It is a well known fact that one of the most important symmetries of string theories, or in general
$\sigma$-models, is target-space, T-duality. There was a common belief among the
experts in this field that the existence of T-duality mainly should depend on the isometry symmerty of the original
target manifold such that, depending on the kind of isometry it 
was called abelian or non-abelian T-duality (for a review see \cite{Giv}). Of course
there were some difficulties in the non-abelian case, for example it was not possible
to obtain the original model from the dual one, because the latter might not have isometry \cite{dela}.
Indeed Klimcik and Severa, by introducing the Poisson-Lie T-duality in their pioneering work \cite{KS1},
could remove the requirement of isometry in obtaining the dual model.
Hence they could remove the above mentioned difficulties in T-duality
investigation. Indeed, Poisson-Lie duality does not require
the isometry in the original target manifold; the integrability
of the Noether's current associated with the action of group $G$ on the target
manifold is good enough to have this symmetry.
In other words, the components of the Noether's current play the role of
flat connection, that is, they satisfy Maurer-Cartan equations with group
structure of $\tilde{G}$ (with the same dimension as $G$) \cite{KS1}; such that 
$G$ and $\tilde{G}$ have Poisson-Lie structure and their Lie algebras form a
bialgebra \cite{D,AM}.
Classically the dual models are equivalent, since one can obtain one 
from the other through a canonical transformation in their
respective phase space. This canonical equivalence has already been shown,
both in Poisson-Lie T-duality \cite{KS1,KS2,Sf1} and abelian
or non-abelian T-duality \cite{C}.
Actually, the quantum equivalence of both abelian and non-abelian dual models
have already been investigated. This equivalence can be shown by obtaining a
general relation between the Weyl-anomaly coefficients (and $\beta$-functions)
of the original model and those of its dual \cite{H,B}. But the quantum equivalence of the
Poisson-Lie dual models is still a challenging problem; the quantum
equivalence has been shown only in some special examples \cite{B,Sf2}.
Therefore, in order to understand the quantum features of the Poisson-Lie
T-duality we have a long way ahead. Hence, we need to investigate
too many $\sigma$-models with Poisson-Lie symmetry, since so far few examples with Poisson-Lie
symmetry have been obtained \cite{KS1,B,Sf2,AKT}. This suggests 
us trying to find further examples with Poisson-Lie symmetry. In this paper,
we find explicitly all possible dual algebras of all three dimensional
real Lie algebras (Bianchi algebras). Then by introducing a non-degenerate adjoint
invariant inner product over the Lie algebras of Drinfeld doubles, we find many 
bialgebras, where we have associated a pair of Poisson-Lie dual $\sigma$-models
to each of these bialgebras which contain all examples mentioned above and many other examples. This
paper is organized as follows. 
In order the paper to be self-contained and also to fix the notations, we give
a brief review of the Poisson-Lie T-duality in section 2. In section 3 we
obtain all possible dual algebras of each of Bianchi algebras, where the 
bialgebras thus obtained have been listed in Table 2. Then, we introduce
a non-degenerate adjoint invariant inner product over (six dimensional) Lie
algebras of Drinfeld doubles, such that each of bialgebras are isotropic with
respect to this inner product.
In section 4 some of the general formulas related to the calculation of
coupling matrices of $\sigma$-models, associated with the bialgebras of Table
2, have been constructed.
Actually these informations help us to write a pair of Poisson-Lie dual
$\sigma$-models associated with each bialgebra of Table 2. At the end of
this section we give two simple examples. Finally, the paper ends
with a brief conclusion and two appendices.

\section{\bf Poisson-Lie T-duality} 

$\; \; \; \;$In this section we review briefly the Poisson-Lie T-duality. 
According to \cite{KS1} the Poisson-Lie duality based on the concepts of Drinfeld 
double and Manin triple. Drinfeld double $D$ is a
Lie group where its Lie algebra $\cal D$ can be decomposed (as a vector space) into
direct sum of two Lie subalgebras ${\bf g}$ and ${\bf \tilde{g}}$, such that this is
maximal isotropic with respect to a non-degenerate invariant bilinear form $< , >$ over
$\cal D$. The doublet $({\bf g}, {\bf \tilde{g}})$ and triplet $({\cal D}, {\bf g}, {\bf \tilde{g}})$
are called bialgebra and Manin triple respectively \cite{D,AM}.
Actually taking the sets $\{X_i\}$ and $\{\tilde{X}^i\}$ as the bases of the Lie
algebras ${\bf g}$ and ${\bf \tilde{g}}$, respectively, we have:
$$
[X_i , X_j] = {f_{ij}}^k X_k,\hspace{20mm} [\tilde{X}^i ,\tilde{ X}^j] ={{\tilde{f}}^{ij}}_{\; \; \: k} {\tilde{X}^k}, \\
$$
\begin{equation}
[X_i , \tilde{X}^j] ={\tilde{f}^{jk}}_{\; \; \; \:i} X_k +  {f_{ki}}^j \tilde{X}^k.
\label{1}
\end{equation}
The isotropy of the subalgebras with respect to bilinear form means that:
\begin{equation}
<X_i , X_j> = <\tilde{X}^i , \tilde{X}^j> = 0, \hspace{16mm} <X_i , \tilde{X}^j> = {\delta_i}^j .
\label{2}
\end{equation}
Also, the Jacobi identity of Lie algebra $\cal D$ imposes the following relations over the
structure constants of Lie algebras ${\bf g}$ and ${\bf \tilde{g}}$ \cite{KS1,AM}:
\begin{equation}
{f_{mk}}^i{\tilde{f}^{jm}}_{\; \; \; \; \; l} - {f_{ml}}^i{\tilde{f}^{jm}}_{\; \; \; \; \; k} - {f_{mk}}^j{\tilde{f}^{im}}_{\; \; \; \; \; l} + {f_{ml}}^j{\tilde{f}^{im}}_{\; \; \; \; \; k} = {f_{kl}}^m{\tilde{f}^{ij}}_{\; \; \; m} .
\label{3}
\end{equation}
As we will see in the next section the above relations put rather strong constraint on the
structure constants of subalgebras. Hence, obtaining of the bialgebra and Manin triple becomes rather a tough job.
In order to define $\sigma$ models with Poisson-Lie duality symmetry, we need to
consider the following relations \cite{KS1}: 
$$
g^{-1} X_i g ={a(g)_i}^j X_j,\hspace{10mm} g^{-1} \tilde{X}^i g = {b(g)}^{ij} X_j + {d(g)^i}_j \tilde{X}^j,
$$
\begin{equation}
\Pi(g) = b(g) a^{-1}(g),
\label{4}
\end{equation}
with $g$ and $\tilde{g}$ as elements of Lie groups $G$ and $\tilde{G}$ associated 
with Lie algebras ${\bf g}$ and ${\bf \tilde{g}}$, respectively. The invariance of inner product 
with respect to adjoint action of group together with (\ref{2}) and (\ref{4}) requires 
the above matrices to possess the following properties:
\begin{equation}
d(g) = a^{-t}(g), \hspace{10mm}a^{-1}(g) = a(g^{-1}), \hspace{10mm} b^t(g) = b(g^{-1}), \hspace{10mm}\Pi^t(g) = -\Pi(g).
\end{equation}
The matrices $\tilde{a}(\tilde{g})$, $\tilde{b}(\tilde{g})$ and $\tilde{\Pi}(\tilde{g})$ 
associated with group $\tilde{G}$ can be defined in a similar way. Now, we can define below the  
$\sigma$-model with d-dimensional target manifold $M$, where the group $G$ acts freely 
on it \cite{KS1,KS2,Sf1}:
$$ 
S = -\frac{1}{2}\int\!d\xi^{+}d\xi^{-}[E_{ij}(\partial_{+}g g^{-1})^i (\partial_{-}g g^{-1})^j + {\Phi^{(1)}}_{i \alpha}(\partial_{+}g g^{-1})^i {\partial_{-}}y^{\alpha} +  {\Phi^{(2)}}_{\alpha i}{\partial_{+}}y^{\alpha}(\partial_{-}g g^{-1})^i 
$$
\begin{equation}
\hspace{20mm} + \Phi_{\alpha \beta}{\partial_{+}}y^{\alpha}{\partial_{-}}y^{\beta}],
\label{6}
\end{equation}
where the coupling matrices are:
$$
E = ({E_0}^{-1} + \Pi)^{-1},\hspace{10mm}\Phi^{(1)} = E {E_0}^{-1} F^{(1)},\hspace{10mm}\Phi^{(2)} = F^{(2)} {E_0}^{-1} E,
$$
\begin{equation}
\Phi = F - F^{(2)} \Pi E {E_0}^{-1} F^{(1)}.
\end{equation}
The $y^{\alpha}$ with $(\alpha = 1,...,d-dimG)$ are coordinates of $M/G$ manifold.
The matrices ${E_0}^{-1}, F^{(1)}, F^{(2)}$ and $F$ are arbitrary functions of $y^{\alpha}$ only.
Similarly, the coupling matrices of the dual $\sigma$-model can be written as \cite{KS2,Sf1}:
$$
\tilde{E} = (E_0 + \tilde{\Pi})^{-1},\hspace{10mm}\tilde{\Phi}^{(1)} = \tilde{E} F^{(1)},\hspace{10mm}\tilde{\Phi}^{(2)} = -F^{(2)} \tilde{E},
$$
\begin{equation}
\tilde{\Phi} = F - F^{(2)} \tilde{E} F^{(1)}.
\end{equation}
The target space of dual model is d-dimensional manifold $\tilde{M}$ with the
group $\tilde{G}$ acting freely on it. The corresponding dual action can be written as:
$$ 
\tilde{S} = -\frac{1}{2}\int\!d\xi^{+}d\xi^{-}[\tilde{E}^{ij}(\partial_{+}\tilde{g} \tilde{g}^{-1})_i (\partial_{-}\tilde{g} \tilde{g}^{-1})_j +
{{\tilde{\Phi}}^{(1) i}}_{\: \: \: \: \: \: \: \alpha}(\partial_{+}\tilde{g} \tilde{g}^{-1})_i {\partial_{-}}y^{\alpha} + {{\tilde{\Phi}}^{(2)\: \: i}}_{\: \; \; \; \alpha}{\partial_{+}}y^{\alpha}(\partial_{-}\tilde{g} \tilde{g}^{-1})_i
$$
\begin{equation}
\hspace{20mm} + \tilde{\Phi}_{\alpha \beta}{\partial_{+}}y^{\alpha}{\partial_{-}}y^{\beta}].
\label{9}
\end{equation}
The actions (6) and (9) correspond to Poisson-Lie dual $\sigma$ models \cite{KS1} 
\footnote{Actually the Noether's current associated with the action of $G(\tilde{G})$ on $M(\tilde{M})$ satisfy
the Maurer-Cartan equation with structure constants of dual Lie group $\tilde{G}(G)$.}.
Notice that if the group $G(\tilde{G})$ besides having free action on $M(\tilde{M})$, acts transitively over it,
then the corresponding manifolds $M(\tilde{M})$ will be the same as the groups $G(\tilde{G})$.
In this case only the first term appears in the actions (6) and (9).
Also if the group $G$ becomes the isometry group of the manifold $M$ with dual abelian group $\tilde{G}$,
then we get the standard nonabelian duality \cite{dela}.

\section{\bf Bianchi Bialgebras}

$\; \; \; \;$In this section, we use (\ref{3}) to obtian the dual Lie algebras ${\bf \tilde{g}}$ of
a given 3-dimensional real Lie algebras ${\bf g}$. Then, by introducing a nondegenerate adjoint
invariant bilinear form we get the corresponding Manin triples. According to Behr's classification \cite{Be,Shep} 
of 3-dimensional Bianchi Lie algebras \cite{T}, the commutation relation of these algebras can be generally written
as:
$$
[X_1 , X_2] = -aX_2 + n_3X_3, \hspace{20mm}[X_2 , X_3] = n_1X_1,
$$
\begin{equation}
[X_3 , X_1] = n_2X_2 + aX_3,
\label{10}
\end{equation}
where the structure constants are given in Table 1 \footnote{For three algebras $III$, $VI_a$ and $VII_a$,
with nonzero parameters $a$, $n_2$ and $n_3$, rescaling of the basis yields the ratio 
$\frac{a^2}{n_2n_3}$ constant. For other algebras one can choose a basis with structure
constants $\pm 1$, as in Table 1.}.

\begin{center}
\hspace{3mm}{\bf Table 1} : 
\hspace{.5mm}Bianchi classification of three dimensional Lie algebras.
\begin{tabular}{|c|c|c|c|c|} \hline\hline
Type                       & $a$ & $n_1$ & $n_2$ & $n_3$  \\ \hline
$I$                        & 0   & 0     & 0     & 0      \\     
$II$                       & 0   & 1     & 0     & 0      \\    
$VII_0$                    & 0   & 1     & 1     & 0      \\   
$VI_0$                     & 0   & 1     & -1    & 0      \\
$IX$                       & 0   & 1     & 1     & 1      \\      
$VIII$                     & 0   & 1     & 1     & -1     \\
$V$                        & 1   & 0     & 0     & 0      \\ 
$IV$                       & 1   & 0     & 0     & 1      \\
$VII_a$                    & $a$ & 0     & 1     & 1      \\
$
\left. \begin{array}{ll}
III  & (a=1) \\
VI_a & (a\neq1)
\end{array} \right\}
$
                           & $a$ & 0     & 1     & -1     \\  \hline
\end{tabular}
\end{center}

For all Lie algebras of Table 1 we have $an_1=0$. The algebras with $a=0$ are called class $A$ 
while those with nonvanishing $a$ are called class $B$. Now we can determine the dual
Lie algebra $\bf {\tilde{g}}$, by using the structure constant of Lie algebra $\bf g$
given in (\ref{10}). Due to the antisymmetricity of (3) with respect to the set of
indices $(i , j)$ and $(k , l)$, these relations, after some algebraic calculations and with the help of (10),
lead to the following nine equations
$$
n_1{\tilde{f}^{23}}_{\; \; \; \; 1} - n_2{\tilde{f}^{13}}_{\; \; \; \; 2} = n_3{\tilde{f}^{12}}_{\; \; \; \; 3},
$$
$$
n_2{\tilde{f}^{13}}_{\; \; \; \; 3} = n_2{\tilde{f}^{12}}_{\; \; \; \; 2}, \hspace{5mm}n_3{\tilde{f}^{13}}_{\; \; \; \; 3} = n_3{\tilde{f}^{12}}_{\; \; \; \; 2}, \hspace{5mm}n_1{\tilde{f}^{23}}_{\; \; \; \; 3} = -n_1{\tilde{f}^{12}}_{\; \; \; \; 1}, \hspace{5mm}n_1{\tilde{f}^{23}}_{\; \; \; \; 2} = n_1{\tilde{f}^{13}}_{\; \; \; \; 1},
$$
$$
-n_1{\tilde{f}^{32}}_{\; \; \; \; 1}+ n_3{\tilde{f}^{12}}_{\; \; \; \; 3} = -n_2{\tilde{f}^{13}}_{\; \; \; \; 2} ,
$$
$$
a({\tilde{f}^{32}}_{\; \; \; \; 2} + {\tilde{f}^{31}}_{\; \; \; \; 1}) = n_3({\tilde{f}^{12}}_{\; \; \; \; 1} + {\tilde{f}^{23}}_{\; \; \; \; 3}) ,
$$
$$
a({\tilde{f}^{32}}_{\; \; \; \; 3} + {\tilde{f}^{12}}_{\; \; \; \; 1}) = n_2({\tilde{f}^{13}}_{\; \; \; \; 1} - {\tilde{f}^{23}}_{\; \; \; \; 2}) ,
$$
\begin{equation}
-a{\tilde{f}^{31}}_{\; \; \; \; 3} + n_2{\tilde{f}^{31}}_{\; \; \; \; 2} - n_3{\tilde{f}^{21}}_{\; \; \; \; 3} - a{\tilde{f}^{21}}_{\; \; \; \; 2} = n_1{\tilde{f}^{23}}_{\; \; \; \; 1} .
\end{equation}
In the above equations some coeffecients can be eliminated, but the reason for their presence
is due to their vanishing for some algebras. Now, in order to determine the structure
constants of dual Lie algebras $\bf \tilde{g}$ associated with those listed in Table 1,
we could write (11) for each of them. Then, by using the Jacobi identity for $\bf \tilde{g}$,
we can determine ${\tilde{f}^{ij}}_{\; \; \; k}$ \footnote{After using (11) for each
Lie algebra, the number of equations becomes less than nine.}. The algebras $\bf \tilde{g}$ reduces to Lie algebras of
Bianchi type if (a) we use the following
general form of structure constant of a given 3-dimensional Lie algebra:
$$
{\tilde{f}^{ij}}_{\; \; \; k} = \epsilon^{ijk}\tilde{n}_{lk} + {\delta^i}_k\tilde{a}^k - {\delta^j}_k\tilde{a}^i,
$$
with $\tilde{a}^i$ as components of a 3-vector and $\tilde{n}_{ij}$ as elements of symmetric matrix with
the constraint $\tilde{n}_{ij}\tilde{a}^j = 0$; and (b) if we work in a basis that diagonalizes
the matrix $\tilde{n}_{ij}$ with diagonal elements $(\tilde{n}_1, \tilde{n}_2, \tilde{n}_3)$ and
vector $\tilde{a}^k$ with components $(\tilde{a}, 0, 0)$. 
But, instead of performing the above steps, we can choose another method with
rather less camputations and obtain the same results, as follows. 
Similar to ${{f}_{ij}}^k$ we assume that the ${\tilde{f}^{ij}}_{\; \; \; k}$ have the following form:
$$
[\tilde{X}_1 , \tilde{X}_2] = -\tilde{a}\tilde{X}_2 + \tilde{n}_3\tilde{X}_3, \hspace{20mm}[\tilde{X}_2 , \tilde{X}_3] = \tilde{n}_1\tilde{X}_1,
$$
\begin{equation}
[\tilde{X}_3 , \tilde{X}_1] = \tilde{n}_2\tilde{X}_2 +\tilde{a}\tilde{X}_3.
\label{12}
\end{equation}
Then (11) reduces to
\begin{equation}
n_1\tilde{n}_1 = 0, \hspace{20mm} a\tilde{a} = n_2\tilde{n}_2 = n_3\tilde{n}_3.
\end{equation}
Hence the structure constants $(\tilde{a}, \tilde{n}_1, \tilde{n}_2, \tilde{n}_3)$ of
the dual Lie algebra ${\bf \tilde{g}}$ can be determined in terms of the structure constants
$(a, n_1, n_2, n_3)$ of the Lie algebra ${\bf g}$. \\
$\; \; \; \;$As an example, for ${\bf g} = IX$, from (13) and Table 1 we have: 
$\tilde{n}_1 = \tilde{n}_2 = \tilde{n}_3 = 0$ and $\tilde{a}$ is an arbitrary constant. 
Now, for $\tilde{a} = 0 $ we obtain Bianchi of type $I$ and for $\tilde{a} \neq 0$ we get Bianchi of type $V$ 
if we rescale the basis such that $\tilde{a} = 1$.
Therefore, the Lie algebra $IX$ has two different dual Lie algebras $I$, $V$. These 
bialgebras have already been used in references \cite{B,LV}. As another example we consider
${\bf g} = VII_0$, then we have $\tilde{n}_1 = \tilde{n}_2 = 0$ and $\tilde{a}$
together with $\tilde{n}_3$ become arbitrary. In this case the dual Lie algebra ${\bf \tilde{g}}$
can be one of the following four different Bianchi Lie algebras:\\
\\
1) For $\tilde{n}_3 = \tilde{a} = 0$, we get Bianchi of type $I$.\\
\\
2) For $\tilde{n}_3\neq 0$ and $\tilde{a} = 0$ via rescaling and change of basis, we
get Bianchi of type $II$.\\
\\
3) For $\tilde{n}_3 = 0$ and $\tilde{a}\neq 0$ via rescaling of basis, we get Bianchi of type $V$.\\
\\
4) Finally, for  $\tilde{n}_3\neq 0$ and $\tilde{a}\neq 0$ via rescaling of basis, we get Bianchi of type $IV$.\\
\\
Therefore the Lie algebra $VII_0$ has four different dual Lie algebras $I$, $II$, $V$, $IV$.
We have also determined the possible dual Lie algebras of other Bianchi Lie algebras,
where the results are given in Table 2.

\begin{center}
\hspace{10mm}{\bf Table 2} : 
\hspace{3mm}Bianchi Bialgebras.
\begin{tabular}{|c|c|}  \hline\hline
${\bf g}$(class $A$)       & ${\bf \tilde{g}}$                                            \\ \hline
$I$                        & all Bianchi algebras                                         \\     
$II$                       & all Bianchi algebras except types of $IX, VIII$                 \\    
$VII_0$                    & $I, II, V, IV$                                               \\   
$VI_0$                     & $I, II, V, IV$                                               \\     
$IX$                       & $I, V$                                                       \\      
$VIII$                     & $I, V$                                                       \\ \hline\hline\hline
${\bf g}$(class $B$)       & ${\bf \tilde{g}}$                                            \\ \hline
$V$                        & class $A$                                                    \\     
$IV$                       & $I, II, VII_0, VI_0$ (class $A$ except types of $IX, VIII$)     \\    
$VII_a$                    & $I, II, VII_{\tilde{a}}$ $(\tilde{a} = \frac{1}{a})$         \\   
$III$                      & $I, II, III$                                                 \\     
$VI_a$                     & $I, II, VI_{\tilde{a}}$ $(\tilde{a} = \frac{1}{a})$          \\  \hline    
\end{tabular}
\end{center}

As in Table 2, all Bianchi Lie algebras, except types of $IX$ and $VIII$, have
more than three different dual Lie algebras. Also Bianchi algebras $II, III, VI_a$ and $VII_a$
can be self dual. The number of different bialgebras listed in Table 2, is $28$
\footnote{It is clear that if the pair $(\bf g, \bf \tilde{g})$ form a bialgebra, then,
also, the pair $(\bf \tilde{g}, \bf g)$ will lead to another bialgebra. Therefore the
total number of bialgebras of Table 2 is $56$.}. So far we have been able to obtain all
possible bialgebras associated with Bianchi algebras. As it is mentioned in the previous
section, in order to have a Manin triple associated with these bialgebras, we need
to have a nondegenerate ad-invariant inner product over the algebra ${\cal D} = {\bf g} \oplus {\bf \tilde{g}}$,
such that the algebras $\bf g$ and $\bf \tilde{g}$ become isotropic with respect to it.
In order to obtain this inner product we choose the set 
$\{T_A\} = \{X_1, X_2, X_3, \tilde{X}_1, \tilde{X}_2, \tilde{X}_3\}$ as the basis of
$\cal D$. Notice that we have more than one $\cal D$ Lie algebra, but considering the
algebras ${\bf g}$ and ${\bf \tilde{g}}$ with commutation relation (10) and (12),
respectively, we can give the general form of $\cal D$. \\
$\; \; \; \;$Now, writing the commutation relation of the algebras $\cal D$ as:
$$
[T_A , T_B] = {C_{AB}}^C T_C,
$$
with $A, B = 1, ..., 6$, the structure constants ${C_{AB}}^C$ can be
obtained by using (1), (10) and (12). Then the basis $T_A$ will have the following matrix form 
in the adjoint representation of $\cal D$:

\begin{equation}
{\cal T}_i = \left( \begin{tabular}{c|c}
                    ${\cal X}_i$       & $0$              \\ \hline \\
                    ${\tilde{\cal Y}}_i$ & $-({\cal X}_i)^t$ \\ 
                    \end{tabular} \right),\hspace{10mm}  {\cal T}_{i+3} = \left( \begin{tabular}{c|c}
                                                                          $-({\tilde{\cal X}}^i)^t$ & ${\cal Y}_i$       \\ \hline \\
                                                                          $0$                     & ${\tilde{\cal X}}^i$   \\ 
                                                                          \end{tabular} \right),
\end{equation}
where ${\cal X}_i$, the adjoint representation of basis of Bianchi Lie algebra ${\bf g}$, 
and the antisymmetric matrices ${\cal Y}_i$ are:
$$
{\cal X}_1 = \left( \begin{array}{ccc}
                      0 & 0   & 0     \\
                      0 & a   & -n_3  \\
                      0 & n_2 & a
                    \end{array} \right), {\cal X}_2 = \left( \begin{array}{ccc}
                                                               0    & -a & n_3  \\
                                                               0    & 0  & 0     \\
                                                               -n_1 & 0  & 0
                                                             \end{array} \right), {\cal X}_3 = \left( \begin{array}{ccc}
                                                                                                        0   & -n_2 & -a \\
                                                                                                        n_1 & 0    & 0    \\
                                                                                                        0   & 0    & 0
                                                                                                      \end{array} \right),
$$
\begin{equation}
{\cal Y}_1 = \left( \begin{array}{ccc}
                      0 & 0     & 0       \\
                      0 & 0     & -n_1  \\
                      0 & n_1   & 0
                    \end{array} \right), {\cal Y}_2 = \left( \begin{array}{ccc}
                                                               0   & a & n_2  \\
                                                               -a  & 0 & 0     \\
                                                              -n_2 & 0 & 0
                                                             \end{array} \right), {\cal Y}_3 = \left( \begin{array}{ccc}
                                                                                                        0   & -n_3 & a \\
                                                                                                        n_3 & 0    & 0   \\
                                                                                                        -a  & 0    & 0
                                                                                                      \end{array} \right).
\end{equation}

Similarly, ${\tilde{\cal X}}^i$, adjoint representation of the basis of dual Bianchi Lie
algebras ${\bf \tilde{g}}$, and ${\tilde{\cal Y}}^i$ have the same form as ${\cal X}_i$ and ${\cal Y}_i$) with
the difference that we must replace the set $(a, n_1, n_2, n_3)$ with $(\tilde{a}, \tilde{n}_1, \tilde{n}_2, \tilde{n}_3)$.\\
$\; \; \; \;$Since, in general, Lie algebra $\cal D$ is non-semisimple, the nondegenerate
ad-invariant inner product on it can not be obtained via the trace of bilinear product of those algebras in adjoint representation.
In other words Killing form is degenerate for these algebras. Therefore, in order to
obtain the nondegenerate ad-invariant metric $\Omega_{AB}$ in the above basis, we have
to solve the following equation \cite{NW}:
\begin{equation}
{C_{AB}}^D \Omega_{CD} + {C_{AC}}^D \Omega_{BD} = 0.
\end{equation}
Equation (16) means that the matrices ${\cal T}_A . \Omega$ must be antisymmetric matrices.
Hence we must find symmetric matrix $\Omega$ such that the above relations holds.
Now, in order to determine $\Omega$ for each bialgebra of Table 2, we must write
${\cal T}_A$ explicitly and then, using the antisymmetricity of ${\cal T}_A .\Omega$, 
we find the explicit form of $\Omega$. As an example, for the pair of $(V, IX)$, the form of $\Omega$ is:
$$
\Omega = \left( \begin{array}{cccccc}
               -b & 0  & 0 & k & 0 & 0  \\
               0  & 0  & 0 & 0 & k & -b \\
               0  & 0  & 0 & 0 & b & k  \\
               k  & 0  & 0 & b & 0 & 0  \\
               0  & k  & b & 0 & b & 0  \\
               0  & -b & k & 0 & 0 & b
               \end{array} \right),\hspace{10mm}det\Omega = -(k^2 + b^2)^3.
$$
Now by choosing $k=1$ and $b=0$ we get:
\begin{equation}
\Omega = \left( \begin{tabular}{c|c}
                 0 & I \\ \hline
                 I & 0 \\ 
                 \end{tabular} \right).
\end{equation}
Actually, by using (14) and (15) one can easily show that the symmetric metric (17) can be ad-invariant metric
of all 28 different bialgebras of Table 2, that is
$$
{\cal T}_A . \left( \begin{tabular}{c|c}
                 0 & I \\ \hline
                 I & 0 \\                                                
                  \end{tabular} \right)
$$
is antisymmetric for all bialgebras of Table 2. Hence, we choose $\Omega$ as the required inner
product. It is also interesting to see that the Lie subalgebras of all bialgebras of Table 2
are isotropic with respect to this inner product, that is we have:
$$
<T_A , T_B> = \Omega_{AB},
$$
$$
<X_i , {\tilde{X}}^j> = <T_i , T_{j+3}> = \Omega_{i,j+3} = \delta_{ij},
$$
\begin{equation}
<{\tilde{X}}^i , X_j> = <T_{i+3} , T_j> = \Omega_{i+3,j} = \delta_{ij}.
\end{equation}
Therefore, all bialgebras of Table 2, together with the ad-invariant metric (17), 
form Manin triples. Now we can construct $\sigma$-models with Poisson-Lie symmetries
associated with the Manin triples obtained above. 

\section{\bf $\sigma$-models and Bianchi Lie Groups}

$\; \; \; \;$As mentioned in section 2, in order to construct $\sigma$-models
with Poisson-Lie symmetry associated with bialgerba $({\bf g} , {\bf \tilde{g}})$,
we need to know $a(g)$ and $b(g)$. Indeed for every bialgebras of Table 2, we
can evaluate $a(g)$ and $b(g)$. Hence we will have a $\sigma$-model as well as its dual,
which possess the Poisson-Lie symmetry. In this section we give general formulas
for $a(g)$ and $b(g)$ in terms of the sets of parameters ($a, n_1, n_2, n_3$) and
($\tilde{a}, {\tilde{n}}_1, {\tilde{n}}_2, {\tilde{n}}_3$). Therefore, by appropriate choices
of these parameters, $a(g)$ and $b(g)$ of all bialgebras of Table 2 can be evaluated.
Hence we can construct all $\sigma$-models associated with them. To achieve this
goal we parametrize the group $G$ in the following form:
\begin{equation}
g = e^{\chi_1 X_1} e^{\chi_2 X_2} e^{\chi_3 X_3},
\end{equation}
where $\chi_i$ are coordinates of the group manifold $G$. In order to calculate
$a(g)$ and $b(g)$, we need to calculate expressions such as $e^{-\chi_i X_i} X_j e^{\chi_i X_i}$ and
$e^{-\chi_i X_i} \tilde{X}^j e^{\chi_i X_i}$. To do this, we must use the algebraic
relations (1), (10) and (12). To make the calculation simple, we need to work with
basis of $\cal D$, where we have \footnote{Here, repeated indices do not imply summation.}:
\begin{equation}
e^{-\chi_i X_i} T_A e^{\chi_i X_i} = e^{-\chi_i T_i} T_A e^{\chi_i T_i} = {(e^{\chi_i {\cal T}_i})_A}^B T_B.
\end{equation}
Therefore, we must calculate matrices $e^{\chi_i {\cal T}_i}$, where due to
the particular block diagonal form of ${\cal T}_i$ (14), in general, they have the following
form:
\begin{equation}
e^{\chi_i {\cal T}_i} = \left( \begin{tabular}{c|c}
                                $e^{\chi_i {\cal X}_i}$ & $0$                   \\ \hline                  
                                $B_i$                   & $e^{-\chi_i ({\cal X}_i)^t}$ \\
                                \end{tabular} \right).
\end{equation}
To calculate (21) we need to know the block diagonal form $D_i$ of matrices ${\cal T}_i$
together with matrices $S_i$, then we have:
\begin{equation}
S_i {\cal T}_i S_i^{-1} = D_i, \hspace{20mm}e^{\chi_i {\cal T}_i} = S_i^{-1} e^{\chi_i D_i} S_i .
\end{equation}
Now we put each ${\cal T}_i$ into its block diagonal form separately. For ${\cal T}_1$ we have: 
\begin{equation}
S_1 = \frac{\tilde{n}_1}{2a} \left( \begin{tabular}{c|c}
                                   $N$ & $\frac{2a}{\tilde{n}_1}$I  \\ \hline
                                   I & 0 \\
                                   \end{tabular} \right), \hspace{20mm} D_1 = \left( \begin{tabular}{c|c}
                                                                                              $-({\cal X}_1)^t$ & $0$         \\ \hline
                                                                                              $0$               & ${\cal X}_1$ \\
                                                                                               \end{tabular} \right),
\end{equation}
where
$$
N = \left( \begin{array}{ccc}
            1 & 0  & 0 \\
            0 & 0  & 1 \\
            0 & -1 & 0
            \end{array} \right).
$$
On the other hand, by using the following matrices:
$$
S = \left( \begin{array}{ccc}
            1 & 0                            & 0 \\
            0 & \sqrt{\frac{n_3}{n_2+n_3}}   & \sqrt{\frac{n_3}{n_2+n_3}} \\
            0 & -i\sqrt{\frac{n_2}{n_2+n_3}} & i\sqrt{\frac{n_2}{n_2+n_3}}
            \end{array} \right),\hspace{10mm} D = \left( \begin{array}{ccc}
                                                                      0 & 0                & 0 \\
                                                                      0 & a+i\sqrt{n_2n_3} & 0 \\
                                                                      0 & 0                & a-i\sqrt{n_2n_3}
                                                                      \end{array} \right),
$$
we write the matrix ${\cal X}_1$ in digonal form, then using (22) for these
matrices we calculate $e^{\chi_1 {\cal X}_1}$:
\begin{equation}
e^{\chi_1 {\cal X}_1} = e^{a\chi_1} \left( \begin{array}{ccc}
                                            e^{-a\chi_1} & 0                                   & 0 \\
                                            0            & \cos{\eta_1}                        & -\sqrt{\frac{n_3}{n_2}} \sin{\eta_1} \\
                                            0            & \sqrt{\frac{n_2}{n_3}} \sin{\eta_1} & \cos{\eta_1}
                                            \end{array} \right),
\end{equation}
where $\eta_1 = \sqrt{n_2n_3}\chi_1$. With the help of (22), (23) and (24) and after some algebraic calculations
we have:
\begin{equation}
B_1 = -\frac{\tilde{n}_1}{a} \sinh{a\chi_1} \left( \begin{array}{ccc}
                                            0 & 0                                   & 0 \\
                                            0 & \sqrt{\frac{n_2}{n_3}} \sin{\eta_1} & \cos{\eta_1} \\
                                            0 & -\cos{\eta_1}                       & \sqrt{\frac{n_3}{n_2}} \sin{\eta_1}
                                            \end{array} \right).
\end{equation}
Similarly, for the matrices ${\cal T}_2$ and ${\cal T}_3$, we have calculated the matrices $S_2$ and $S_3$,
which transform them into their Jordan forms $D_2$ and $D_3$ (these matrices are given in appendix A).
Finally, using these Jordan forms together with (22) and considering (21), we obtain the
following expressions for matrices $e^{\chi_2 {\cal T}_2}$ and $e^{\chi_3 {\cal T}_3}$:
\begin{equation}
e^{\chi_2 {\cal X}_2} = \left( \begin{array}{ccc}
                               \cos{\eta_2}                        & -\frac{a}{\sqrt{n_1 n_3}}\sin{\eta_2}  & \sqrt{\frac{n_3}{n_1}}\sin{\eta_2} \\
                                0                                  & 1                                       & 0 \\
                               -\sqrt{\frac{n_1}{n_3}}\sin{\eta_2} & -\frac{a}{n_3}(\cos{\eta_2}-1)         & \cos{\eta_2}
                                \end{array} \right),
\end{equation}
where $\eta_2 = \sqrt{n_1n_3}\chi_2$, and
\begin{equation}
\hspace{-18mm}B_2 = \left( \begin{array}{ccc}
              -\sqrt{\frac{n_1}{n_3}}\chi_2\tilde{n}_2\sin{\eta_2}                                                         & \frac{\xi}{n_3\sqrt{n_1n_3}}\sin{\eta_2} - \frac{a\tilde{n}_2\chi_2}{n_3}\cos{\eta_2}         &  \tilde{n}_2\chi_2\cos{\eta_2} \\
              -\frac{\xi}{n_3\sqrt{n_1n_3}}\sin{\eta_2} + \frac{a\tilde{n}_2\chi_2}{n_3}\cos{\eta_2} & -\frac{2a\xi}{n_1{n_3}^2}(\cos{\eta_2} - 1) - \frac{a^2\tilde{n}_2\chi_2}{n_3\sqrt{n_1n_3}}\sin{\eta_2}           &  \frac{\xi}{n_1n_3}(\cos{\eta_2} - 1) + \frac{a\tilde{n}_2\chi_2}{\sqrt{n_1n_3}}\sin{\eta_2} \\
              -\tilde{n}_2\chi_2\cos{\eta_2}                                                                               & \frac{\xi}{n_1n_3}(\cos{\eta_2} - 1) + \frac{a\tilde{n}_2\chi_2}{\sqrt{n_1n_3}}\sin{\eta_2}   &  -\sqrt{\frac{n_3}{n_1}}\chi_2\tilde{n}_2\sin{\eta_2}             
             \end{array} \right).
\end{equation}
where $\xi = \tilde{a}n_3 + \tilde{n}_2a$.\\
Also
\begin{equation}
e^{\chi_3 {\cal X}_3} = \left( \begin{array}{ccc}
                               \cos{\eta_3}                        & -\sqrt{\frac{n_2}{n_1}}\sin{\eta_3} & -\frac{a}{\sqrt{n_1 n_2}}\sin{\eta_3} \\
                               \sqrt{\frac{n_1}{n_2}}\sin{\eta_3}  & \cos{\eta_3}                        & \frac{a}{n_2}(\cos{\eta_3}-1) \\
                               0                                   & 0                                   & 1                  
                               \end{array} \right),
\end{equation}
where $\eta_3 = \sqrt{n_1n_2}\chi_3$, and
\begin{equation}
\hspace{-18mm}B_3 = \left( \begin{array}{ccc}
              -\sqrt{\frac{n_1}{n_2}}\chi_3\tilde{n}_3\sin{\eta_3}                                                         & -\tilde{n}_3\chi_3\cos{\eta_3}                                                                                     &  \frac{\tilde{\xi}}{n_2\sqrt{n_1n_2}}\sin{\eta_3} - \frac{a\tilde{n}_3\chi_3}{n_2}\cos{\eta_3}         \\  
              \tilde{n}_3\chi_3\cos{\eta_3}                                                                                & -\sqrt{\frac{n_2}{n_1}}\chi_3\tilde{n}_3\sin{\eta_3}                                                               &  -\frac{\tilde{\xi}}{n_1n_2}(\cos{\eta_3} - 1) - \frac{a\tilde{n}_3\chi_3}{\sqrt{n_1n_2}}\sin{\eta_3}  \\
              -\frac{\tilde{\xi}}{n_2\sqrt{n_1n_2}}\sin{\eta_3} + \frac{a\tilde{n}_3\chi_3}{n_2}\cos{\eta_3} & -\frac{\tilde{\xi}}{n_1n_2}(\cos{\eta_3} - 1) - \frac{a\tilde{n}_3\chi_3}{\sqrt{n_1n_2}}\sin{\eta_3} &  -\frac{2a\tilde{\xi}}{n_1{n_2}^2}(\cos{\eta_3} - 1) - \frac{a^2\tilde{n}_3\chi_3}{n_2\sqrt{n_1n_2}}\sin{\eta_3}
              \end{array} \right).
\end{equation}
where $\tilde{\xi} = \tilde{a}n_2 + \tilde{n}_3a$.\\

Now, considering the defenitions of $b(g)$ and $a(g)$ given in (4), in general we have \cite{KS1}:
\begin{equation}
g^{-1} \left( \begin{array}{c}
               T_1 \\ \vdots \\T_6
               \end{array} \right) g = \left( \begin{tabular}{c|c}
                                               $a(g)$ & $0$   \\ \hline
                                               $b(g)$ & $d(g)$ \\
                                               \end{tabular} \right) \left( \begin{array}{c}
                                                                          T_1 \\ \vdots \\T_6
                                                                          \end{array} \right).
\end{equation}
Hence due to (19) and (20) we have:
\begin{equation}
e^{\chi_1{\cal T}_1} e^{\chi_2{\cal T}_2} e^{\chi_3{\cal T}_3} = \left( \begin{tabular}{c|c}
                                                                        $a(g)$ & $0$   \\ \hline
                                                                        $b(g)$ & $d(g)$ \\
                                                                        \end{tabular} \right).
\end{equation}
Finally, by using (21) we get:
$$
a(g) = e^{\chi_1{\cal X}_1} e^{\chi_2{\cal X}_2} e^{\chi_3{\cal X}_3},
$$
\begin{equation}
b(g) = B_1 e^{\chi_2{\cal X}_2} e^{\chi_3{\cal X}_3} + e^{-\chi_1{{\cal X}_1}^t} B_2 e^{\chi_3{\cal X}_3} + e^{-\chi_1{{\cal X}_1}^t} e^{-\chi_2{{\cal X}_2}^t} B_3.
\end{equation}
Indeed by the above mentioned prescription, we can evaluate $\tilde{a}(\tilde{g})$ and
$\tilde{b}(\tilde{g})$. Since the set of $a(g)$ and $b(g)$ associated with bialgebra
$({\bf \tilde{g}} , {\bf g})$ is the same as the set $\tilde{a}(\tilde{g})$ and $\tilde{b}(\tilde{g})$
of bialgebra $({\bf g} , {\bf \tilde{g}})$, therefore in order to write $\tilde{a}(\tilde{g})$ and $\tilde{b}(\tilde{g})$, 
it is enough to replace the parameters $\chi_i$ and $(a, n_1, n_2, n_3)$ with $\tilde{\chi}_i$ and $(\tilde{a}, \tilde{n}_1, \tilde{n}_2, \tilde{n}_3)$ in (31) and (32), respectively.
As an example, in order to determine
$\tilde{a}(\tilde{g})$ and $\tilde{b}(\tilde{g})$ of the pair ($IX , V$), we just need to calculate
$a(g)$ and $b(g)$ of the pair ($V , IX$), provided that we replace the coordinates 
$\chi_i$ with ${\tilde{\chi}^i}$.\\
So far all calculations have been done on the basis of the parametrization (19) for Bianchi Lie groups $G$.
In the case of usual parametrization of each of Bianchi groups \footnote{ For the group $IX$
this parametrization is of Eulerian type, and for other groups the form of parametrization
is $g=e^{\chi_1 X_3}e^{\chi_3 X_2}e^{\chi_2 X_1}$ \cite{T}.}, one only needs to consider the order of 
multiplication of matrices $e^{\chi_i {\cal T}_i}$ in (31), based on the chosen group parametrization,
then the required relations (32) can be obtained for this new representation. \\
Now, knowing the matrices $\Pi(g)$ and $\tilde{\Pi}(\tilde{g})$ we can write the
related $\sigma$-models and their Poisson-Lie dual models. As it has been already mentioned,
we can associate a pair of dual $\sigma$ models to each bialgebra of Table 2.
Of course some of these models have already been studied. For example models related to the bialgebras ($I , \bf g$)
(standard nonabelian duality) with $M$ as four dimensional space-time manifold (Bianchi cosmological models) have already
been studied in details \cite{GV}. Dual $\sigma$-models associated with bialgebra ($IX , V$)
have been investigated in \cite{B,Sf2,LV}. Also in \cite{AKT} the
dual models of bialgebra ($VIII , V$) have been considered. Here in this article we
just express two pair of dual $\sigma$-models associated with bialgebras of Table 2. \\
In the following examples, we assume that $M$ is a four dimensional manifold with 
($\{y^{\alpha}\}=\{t\}$). Also, we assume that $F^{(1)}=F^{(2)}=0$ and $F=1$,
while the group parametrization is the same as (19). Hence the action (6) reads:
\begin{equation}
S = -\frac{1}{2}\int\!d\xi^{+}d\xi^{-}[{({E_0}^{-1} + \Pi)^{-1}}_{ij}(\partial_{+}g g^{-1})^i (\partial_{-}g g^{-1})^j + \partial_{+}t \partial_{-}t],
\end{equation}
and from (9) its dual action is:
\begin{equation} 
\tilde{S} = -\frac{1}{2}\int\!d\xi^{+}d\xi^{-}[{(E_0 + \tilde{\Pi})^{-1}}^{\: \: ij}(\partial_{+}\tilde{g} \tilde{g}^{-1})_i (\partial_{-}\tilde{g} \tilde{g}^{-1})_j + \partial_{+}t \partial_{-}t].
\end{equation}
As an example we first start with bialgebra ($II , II$) which is a typical example of self-dual bialgebras.
Using the informations of Table 1 together with the relations (24-29) we can evaluate all
matrices needed to determine $\Pi(g)$ \footnote{In caculating these matrices we must remove a 
series of ambiguities.}, where these have been given in the appendix B. Finally the matrix $\Pi(g)$ of this
model can be written as:
\begin{equation}
\Pi(g) = B_1 = \left( \begin{array}{ccc}
                       0 & 0      & 0 \\
                       0 & 0      & -\chi_1 \\
                       0 & \chi_1 & 0
                       \end{array} \right).
\end{equation}
Also we have
\begin{equation}
dg.g^{-1} = (d\chi_1 + \chi_2d\chi_3)X_1 + d\chi_2X_2 + d\chi_3X_3.
\end{equation}
Now by considering $E_0=I$, from (33) we have (in world sheet coordinate):
\begin{equation}
S = -\frac{1}{2}\int\!d\sigma d\tau[\partial_\mu\chi_1 \partial^\mu\chi_1 + 2\chi_2\partial_\mu\chi_1 \partial^\mu\chi_3 + 
\frac{1}{1+{\chi_1}^2}\partial_\mu\chi_2 \partial^\mu\chi_2 + ({\chi_2}^2+\frac{1}{1+{\chi_1}^2})\partial_\mu\chi_3 \partial^\mu\chi_3 +
\partial_\mu t \partial^\mu t].
\end{equation}
Actually this is a self-dual model with respect to Poisson-Lie dual transformation.
Indeed if we choose $E_0$ as an arbitrary invertible constant matrix then the dual
$\sigma$-model again will be the same as the original one but we must replace elements of $E_0$
with those of ${E_0}^{-1}$. Actually this particular example is similar to two dimensional
Borelian dual models of reference \cite{KS1}. \\
Another example is related to the bialgebra ($II , V$). Again, after calculating the matrices
(24-29) given in appendix B, $\Pi(g)$ can be written as:
\begin{equation}
\Pi(g) = \left( \begin{array}{ccc}
                 0       & \chi_2 & \chi_3 \\
                 -\chi_2 & 0      & 0 \\
                 -\chi_3 & 0      & 0
                \end{array} \right).
\end{equation}
Here $dg.g^{-1}$ has the same form as (36). Hence for $E_0=I$, from (33) the action of the original model is:
$$
S = -\frac{1}{2}\int\!d\sigma d\tau\{\frac{1}{1+{\chi_2}^2+{\chi_3}^2}[\partial_\mu\chi_1 \partial^\mu\chi_1 + 2\chi_2\partial_\mu\chi_1 \partial^\mu\chi_3 + 
(1+{\chi_3}^2)\partial_\mu\chi_2 \partial^\mu\chi_2
$$
\begin{equation}
\hspace{30mm}+ (1+2{\chi_2}^2)\partial_\mu\chi_3 \partial^\mu\chi_3] + \partial_\mu t \partial^\mu t\}.
\end{equation}
Finally, by calculating matrices (24-29), as given in appendix B, we determine $\tilde{\Pi}(\tilde{g})$:
\begin{equation}
\tilde{\Pi}(\tilde{g}) = e^{-\tilde{\chi}_1} \left( \begin{array}{ccc}
                                            0 & 0                     & 0 \\
                                            0 & 0                     & -\sinh{\tilde{\chi_1}} \\
                                            0 & \sinh{\tilde{\chi_1}} & 0
                                            \end{array} \right).
\end{equation}
On the other hand, we have:
\begin{equation}
d\tilde{g}.{\tilde{g}}^{-1} = d\tilde{\chi}_1 \tilde{X}^1 + e^{-\tilde{\chi}_1}d\tilde{\chi}_2\tilde{X}^2 + e^{-\tilde{\chi}_1}d\tilde{\chi}_3\tilde{X}^3.
\end{equation}
Then from (34) the dual model can be written as:
\begin{equation}
\tilde{S} = -\frac{1}{2}\int\!d\sigma d\tau[\partial_\mu\tilde{\chi}_1 \partial^\mu\tilde{\chi}_1 + \frac{e^{-2\tilde{\chi}_1}}{1+e^{-2\tilde{\chi}_1}{\sinh^2{\tilde{\chi}_1}}}
(\partial_\mu\tilde{\chi}_2 \partial^\mu\tilde{\chi}_2 + \partial_\mu\tilde{\chi}_3 \partial^\mu\tilde{\chi}_3) + 
\partial_\mu t \partial^\mu t]
\end{equation}

\newpage

{\Large \it {\bf Concluding Remarks}}

$\; \; \; \;$We have obtained dual algebras of all Bianchi type algebras. Then by 
introducing a non-degenerate adjoint invariant inner product over (six dimensional) Lie algebra 
of Drinfeld double, we have obtained many real bialgebras listed in Table 2. 
As it has been shown above, we can associate a pair of Poisson-Lie dual $\sigma$-models to 
every pair of bialgebras of Table 2. This point is explained by two examples; 
the rest will apear in furture works. Beside introducing associated $\sigma$-models with
Poisson-Lie dualities, there are other important problems to be studied, 
and we mention some of them in the following. 
\\
Actually determination of the Drinfeld doubles of Table 2 can be very
important for the following reasons. Due to the existence of non-degenerate
adjoint invariant metric on these groups, we can construct some important
physical models such as WZNW models or gauge theories over them (mostly non-compact).
Also one can determine the modular space of each of these Drinfeld doubles.
Below we give the known doubles. Two of these doubles have already been known, namely
the double $(IX , V)$ is $SO(3, 1)$ and $(I, I)$ is $(U(1))^6$.
Our investigations show that the bialgebras of Table 2 are the following types.
The Drinfeld doubles of $(III , I), (III , II), (VI_0 , I), (VII_0 , I)$ and
$(V , I)$ are real forms of the complexification of $H_4 \otimes U(1) \otimes U(1)$. 
Also $(III , III)$ is $U(2) \otimes U(1) \otimes U(1)$.
On the other hand $(V , II)$ and $(VII_a , I)$ are real forms of the
complexification of six dimensional Heisenberg algebra.
Also the doubles $(VII_0 , V), (VI_0 , V), (IX , I)$ and $(VIII , I)$
correspond to the real forms of $ISO(3, c)$. The
doubles $(II , II)$ and $(VIII , V)$ are real forms of $SO(4, c)$.
Finally doubles $(IV , I), (IV , II), (VII_a , II)$ and $(VI_a , II)$ are solvable.
The exact and thorough characterization of these doubles and those mentioned in Table 2
are under further investigation.
\\
On the other hand, by using the information of Table 2 together
with the prescription of references \cite{P1} and \cite{P2} one can study the Poisson-Lie
T-duality in N=2 superconformal WZNW models related to the bialgebras of Table 2, 
for both classical and quantum cases, similar to \cite{P1} and \cite{P2}. Furthermore for
some bialgebras of Table 2 one can extend these studies to N=4 superconformal WZNW models.
\\                                                     
Also, because of canonical transformation nature of Poisson-Lie T-duality, it
would be interesting to find the generating function of these canonical transformations
for each bialgebra of Table 2 \cite{Sf1}; similar to the work done in \cite{Sf2} for ($IX , V$).
Notice that here the algebra $\cal D$ can have more than one decomposition such
as $\cal D= ({\bf g}\oplus{\bf \tilde{g}})$ and $\cal D= ({\bf k}\oplus{\bf \tilde{k}})$.
Hence it is possible to find canonical transformation which relates the $\sigma$-model with
group $G$ to the one with group $K$ \cite{KS2}. On the other hand, one can research
the commutativity of the renormalisation flow with Poisson-Lie T-duality, by explicit
computing of the beta functions and Weyl anomaly coefficients at the 1-loop level
for these $\sigma$-models and their duals. Of course this work has already been done in
\cite{B} for ($IX , V$). Finally, there is a possibility that some of these models have 
application in string cosmology \cite{Gas}; then the Poisson-Lie T-duality will play a 
key role in interelating different cosmological models and their solutions. \\
\\            

\vskip8pt\noindent

{\large{\bf Acknowledgment}}
We wish to thank S.K.A. Seyed Yagoobi for carefully reading the 
article and for his constructive comments. Also we are very grateful to S. E. Parkhomenko
for his useful comments about Poisson-Lie T-duality in SCWZNW models.

\vskip8pt\noindent

{\bf Appendix A}

$\; \; \; \;$Here in this appendix we give the required  similarity transformations for obtaining
Jordan forms of the matrices ${\cal T}_2$ and ${\cal T}_3$.
The similarity transformation related to the matrix ${\cal T}_2$ and its Jordan
form are:
$$
S_2 = \left( \begin{array}{cccccc}
               0                & 1                                   & -\frac{\xi}{n_1}         & 0             & n_3 & a     \\
               0                & 1                                   & 0                        & 0             & 0   & 0     \\
               0                & -i\frac{\xi}{n_3}                   & 0                        & \sqrt{n_1n_3} & 0   & -in_1 \\
               \sqrt{n_1n_3}    & ia                                  & -in_3                    & 0             & 0   & 0     \\
               0                & i\frac{\xi}{n_3}                    & 0                        & \sqrt{n_1n_3} & 0   & in_1  \\
              -i\tilde{n}_2n_1  & -a\tilde{n}_2\sqrt{\frac{n_1}{n_3}} & \tilde{n}_2\sqrt{n_1n_3} & 0             & 0   & 0
               \end{array} \right),\hspace{10mm}D_2 = \left( \begin{array}{c|c|c}
                                                                       0 & 0   & 0 \\ \hline 
                                                                       0 & A_1 & 0 \\ \hline
                                                                       0 & 0   & A_2
                                                                       \end{array} \right),
$$
where:
$$
A_1 = \left( \begin{array}{cc}                                          
             i\sqrt{n_1n_3} & i\tilde{n}_2\sqrt{\frac{n_1}{n_3}} \\
             0              & i\sqrt{n_1n_3}
              \end{array} \right), \hspace{20mm}A_2 = \left( \begin{array}{cc}
                                                                          -i\sqrt{n_1n_3} & 1   \\

                                                                           0              & -i\sqrt{n_1n_3}
                                                                           \end{array} \right).
$$
Similarly, the similarity transformation related to the matrix ${\cal T}_3$ and
its Jordan form are:
$$
S_3 = \left( \begin{array}{cccccc}
               0                & -\frac{\tilde{\xi}}{n_1}         & 0                                   & 0             & a     & -n_2  \\
               0                & 0                                & 1                                   & 0             & 0     & 0     \\
               0                & 0                                & -i\frac{\tilde{\xi}}{n_2}           & \sqrt{n_1n_2} & in_1  & 0     \\
               \sqrt{n_1n_2}    & in_2                             & ia                                  & 0             & 0     & 0     \\
               0                & 0                                & i\frac{\tilde{\xi}}{n_2}            & \sqrt{n_1n_2} & -in_1 & 0     \\
              -i\tilde{n}_3n_1  & -\tilde{n}_3\sqrt{n_1n_2}        & -a\tilde{n}_3\sqrt{\frac{n_1}{n_2}} & 0             & 0     & 0
               \end{array} \right),\hspace{10mm}D_3 = \left( \begin{array}{c|c|c}
                                                                       0 & 0   & 0 \\ \hline 
                                                                       0 & A_3 & 0 \\ \hline
                                                                       0 & 0   & A_4
                                                                       \end{array} \right),
$$
where $A_3$ and $A_4$ are similar to $A_1$ and $A_2$
except that we must replace ($n_3$, $\tilde{n}_2$) by
($n_2$, $\tilde{n}_3$), respectively.\\
\\

{\bf Appendix B}

$\; \; \; \;$In this appendix we give coupling matrices of bialgebras of the 
examples quoted at the end of section 4.
For bialgebra ($II , II$) we have:
$$
e^{\chi_1 {\cal X}_1} = I,\hspace{10mm}B_1 = \left( \begin{array}{ccc}
                                                                0 & 0 & 0 \\
                                                                0 & 0 & -\chi_1 \\
                                                                0 & \chi_1 & 0
                                                                \end{array} \right), \hspace{10mm} B_2 = B_3 = 0 ,
$$
$$
e^{\chi_2 {\cal X}_2} = \left( \begin{array}{ccc}
                            1       & 0 & 0 \\
                            0       & 1 & 0 \\
                            -\chi_2 & 0 & 1
                            \end{array} \right),\hspace{20mm} e^{\chi_3 {\cal X}_3} = \left( \begin{array}{ccc}
                                                                                                          1       & 0 & 0 \\
                                                                                                          \chi_3  & 1 & 0 \\
                                                                                                          0       & 0 & 1
                                                                                                          \end{array} \right).
$$
Similarly, for bialgebra ($II , V$) we have:
$$
e^{\chi_1 {\cal X}_1} = I,\hspace{10mm}B_1 = 0,\hspace{10mm}B_2 = \left( \begin{array}{ccc}
                                                                                                0       & \chi_2 & 0 \\
                                                                                                -\chi_2 & 0      & 0 \\
                                                                                                0       & 0      & 0
                                                                                                \end{array} \right),\hspace{10mm}B_3 = \left( \begin{array}{ccc}
                                                                                                                                                         0       & 0 & \chi_3 \\
                                                                                                                                                         0       & 0 & 0 \\
                                                                                                                                                         -\chi_3 & 0 & 0
                                                                                                                                                         \end{array} \right),
$$
where the matrices $e^{\chi_2 {\cal X}_2}$ and $e^{\chi_3 {\cal X}_3}$ have
the same form as the matrices for bialgerba ($II , II$), as mentioned above.
Finally the matrices related to the dual model of bialgebra ($II , V$) are:
$$
e^{\tilde{\chi}_1 \tilde{\cal X}^1} = e^{\tilde{\chi}_1} I,\hspace{10mm}\tilde{B}_1 = \left( \begin{array}{ccc}
                                                                                                             0 & 0                     & 0 \\
                                                                                                             0 & 0                     & -\sinh{\tilde{\chi}_1} \\
                                                                                                             0 & \sinh{\tilde{\chi}_1} & 0
                                                                                                             \end{array} \right),\hspace{10mm} \tilde{B}_2 = \tilde{B}_3 = 0 ,
$$
$$
e^{\tilde{\chi}_2 \tilde{\cal X}^2} = \left( \begin{array}{ccc}
                                              1 & -\tilde{\chi}_2 & 0 \\
                                              0 & 1               & 0 \\
                                              0 & 0               & 1
                                              \end{array} \right),\hspace{20mm} e^{\tilde{\chi}_3 \tilde{\cal X}^3} = \left( \begin{array}{ccc}
                                                                                                                                             1 & 0 & -\tilde{\chi}_3 \\
                                                                                                                                             0 & 1 & 0 \\
                                                                                                                                             0 & 0 & 1
                                                                                                                                             \end{array} \right).
$$                                                                                                          
\\


\begin{thebibliography}{99}
\bibitem{Giv}A. Giveon, M. Porrati and E. Rabinovici, Phys. Rep {\bf 244} (1994) 77; 
E. Alvarez, L. Alvarez-Gaume and Y. Lozano, Nucl. Phys (Proc. Suppl){\bf 41} (1995) 1.
\bibitem{dela}X. C. dela Ossa and F. Quevedo, Nucl. Phys. B {\bf 403} (1993) 377; A. Giveon and M. Rocek, Nucl. Phys. B {\bf 421} (1994) 173;
E. Alvarez, L. Alvarez-Gaume, J. L. F. Barbon and Y. Lozano, Nucl. Phys. B {\bf 415} (1994) 71;
K. Sfetsos, Phys. Rev. D {\bf 50} (1994) 2784.
\bibitem{KS1}C. Klimcik and P. Severa, Phys. Lett. B {\bf 351}(1995) 455; 
C. Klimcik, Nucl. Phys.(Proc. Suppl.)B {\bf 46} (1996) 116, he-th/9509095.
\bibitem{D}V. G. Drinfeld. {\it "Quantum Groups"}. In; Proc. ICM, MSRI, Berkeley, 1986. p. 798.
\bibitem{AM}A. Yu. Alekseev and A. Z. Malkin, Commun. Math. Phys. {\bf 162} (1994) 147.
\bibitem{KS2}C. Kilmcik and P. Severa, Phys. Lett. B {\bf 372}(1996) 65, hep-th/9512040.
\bibitem{Sf1}K. Sfetsos, Nucl. Phys. B {\bf 517} (1998) 549, hep-th/9710163.
\bibitem{C}T. Curtright and C. Zachos, Phys. Rew. D {\bf 49} (1994) 5408; E. Alvarez, L. Alvarez-Gaume and Y. Lozano, Phys. Lett. B {\bf 336} (1994) 183;
Y. Lozano, Phys. Lett. B {\bf 355} (1995) 165; O. Alvarez and C-H. Liu, Commun. Math. Phys. {\bf 179} (1996) 185.
\bibitem{H}P. E. Haagensen and K. Olsen, Nucl. Phys. B {\bf 504} (1997) 326, hep-th/9704157.
\bibitem{B}J. Balog, P. Forgacs, N. Mohammedi, L. Palla and J. Schnittger, Nucl. Phys. B {\bf 535} 461, hep-th/9806068.
\bibitem{Sf2}K. Sfetsos, Phys. Lett. B {\bf 432} (1998) 365, hep-th/9803019.
\bibitem{AKT}A. Yu. Alekseev, C. Klimcik and A. A. Tseytlin, Nucl. Phys. B {\bf 458} (1996) 430, hep-th/9509123.
\bibitem{Be}F. B. Estabrook, H. D. Wahlquist and C. G. Behr, J. Math. Phys. {\bf 9} (1968) 497;
G. F. R. Ellis and M. A. H. MacCallum, Commun. Math. Phys. {\bf 12} (1969) 108.
\bibitem{Shep}M. Ryan and L. Shepley, {\it "Homogeneous Relativistic Cosmologies"}, Princeton Univ. Press, Princeton, 1975;
M. A. H. MacCallum, {\it "Anisotropic and Inhomogeneous Relativistic Cosmologies"}, in {\it "The Early Universe: Reprints"} by
E. W. Kolb and M. S. Turner, Addison-Wisley. N.Y. 1990; L. D. Landau and E. M. Lifshitz, {\it "The Classical Theory of Fields"}, Pergamon Press, 1987.
\bibitem{T}A. H. Taub, Ann. Math. {\bf 53} (1951) 472.
\bibitem{LV}M. A. Lledo and V. S. Varadarajan, Lett. Math. Phys {\bf 45} (1998) 247, hep-th/9803175; 
G. Marmo, A. Simoni and A. Stern, Int. J. Mod. Phys. A {\bf 10} (1995) 99.
\bibitem{NW}D. Cangemi and R. Jackiw. Phys. Rev. Lett. {\bf 69} (1992) 233; Ann. Phys. {\bf 225} (1993) 229;  
C. R. Nappi and E. Witten, Phys. Rew. Lett. {\bf 71} (1993) 3751.
\bibitem{GV}M. Gasperini, R. Ricci and G. Veneziano, Phys. Lett. B {\bf 319} (1993) 438; S. Elitzur, A. Giveon, E. Rabinovici, A. Schwimmer and G. Venaziano,
Nucl. Phys. B {\bf 435} (1995) 147; E. Alvarez, L. Alvarez-Gaume and Y. Lozano, Nucl. Phys. B {\bf 424} (1994) 155; 
E. Tyurin, Phys. Lett. B {\bf 348} (1995) 386.
\bibitem{Gas}See the Home-Page, http://www.to.infn.it/$\sim $gasperin. 
\bibitem{P1}S. E. Parkhomenko, JETP Lett. {\bf 64} (1996) 877; Mod. Phys. Lett. A {\bf 12} (1997) 3091; Nucl. Phys. B {\bf 510} (1998) 623.
\bibitem{P2}S. E. Parkhomenko, "On the Quantum Poisson-Lie T-duality and Mirror Symmetry" hep-th/9812048.

\end{thebibliography}
\end{document}